\documentclass[prl,twocolumn,showpacs]{revtex4}
\usepackage{amsmath, amsgen,amstext,amsbsy,amsopn,amsthm, amssymb}
\usepackage[dvips]{graphicx}

\theoremstyle{definition}

\usepackage{color}

\def\*{{\phantom *}}
\begin{document}
\title{Quasi-one- and quasi-two-dimensional perfect Bose gas: the second critical density and generalised condensation}

\author{Mathieu Beau}
\author{Valentin A. Zagrebnov}
\affiliation{Universit\'e de la M\'editerran\'ee and Centre de Physique
Th\'eorique - UMR 6207
\\ Luminy - Case 907,
13288 Marseille, Cedex 09, France}

\date{February 15, 2010}

\begin{abstract}
\noindent In this letter we discuss a relevance of the 3D Perfect Bose gas (PBG) condensation
in extremely elongated
vessels for the study of anisotropic condensate coherence and the "quasi-condensate".
To this end we analyze the case of exponentially anisotropic (\textit{van den Berg}) boxes,
when there are
\textit{two} critical densities $\rho_{c} < \rho_{m}$ for a generalised Bose-Einstein Condensation
(BEC). Here
$\rho_{c}$ is the standard critical density for the PBG. We consider three examples of anisotropic
geometry: slabs,
squared beams and "cigars" to demonstrate that the "quasi-condensate" which exists in domain
$\rho_{c} < \rho < \rho_{m}$
is in fact the van den Berg-Lewis-Pul\'{e} generalised condensation (vdBLP-GC) of the type III
with no macroscopic
occupation of any mode.

We show that for the slab geometry the second critical density $\rho_{m}$ is a threshold between
\emph{quasi-two-dimensional} (\emph{quasi-2D}) condensate and the \emph{three dimensional}
(\emph{3D}) regime
when there is a coexistence of the "quasi-condensate" with the standard one-mode BEC.
 On the other hand,
in the case of squared beams and "cigars" geometries critical density $\rho_{m}$ separates
\emph{quasi-1D}
and \emph{3D} regimes.
We calculate the value of difference between $\rho_{c}$, $\rho_{m}$ (and between corresponding
critical temperatures
$T_{m}$, $T_{c}$) to show that observed space anisotropy of the condensate coherence can be
described by a critical
exponent $\gamma(T)$ related to the anisotropic ODLRO.
We compare our calculations with physical results for extremely elongated traps that manifest
"quasi-condensate".

\end{abstract}


\maketitle
\noindent 1.One can rigorously show that there is no a \textit{conventional} Bose-Einstein condensation
(BEC) in the
one- ($1D$) and two-dimensional ($2D$) boson systems or in the three-dimensional squared beams
(cylinders) and
slabs (films). For interacting Bose-gas it results from the Bogoliubov-Hohenberg theorem \cite{Bog62},
\cite{Hoh},
based on a non-trivial Bogoliubov inequality, see e.g. \cite{BoMa}. For the perfect Bose-gas this result
is much easier, since it follows from the explicit analysis of the occupation number density in
one-particle eigenstates. A common point is the Bogoliubov $1/q^2$-theorem \cite{Bog62},
\cite{Bog-Selv2}, \cite{Bog-v6-II}, which implies destruction of the macroscopic occupation of the ground-state
by thermal fluctuations.

Renewed interest to eventual possibility of the "condensate" in the quasi-one-, or -two-dimensional
(quasi-$1D$ or -$2D$) boson gases (i.e., in cigar-shaped systems or slabs) is motivated by recent experimental data
indicating the existence of so-called "quasi-condensate" in anisotropic traps \cite{Gerbier}-\cite{Petrov1} and
BKT crossover \cite{Dalibard}.

The aim of this letter is \textit{twofold}. First we show that a natural modeling of slabs by
highly anisotropic $3D$-cuboid implies in the thermodynamic limit the van den Berg-Lewis-Pul\'{e}
\textit{generalised
condensation} (vdBLP-GC) \cite{vdBLP} of the Perfect Bose-Gas (PBG) for densities larger than the
\textit{first}, i.e.
the standard critical $\rho_{c}(\beta)$ for the inverse temperature $\beta = 1/(k_{B}T)$. Notice,
that a special case
of this (induced by the geometry) condensation was pointed out for the first time by Casimir \cite{Cas},
although the
theoretical concept and the name are due to Girardeau \cite{Gir}. So, for the PBG the "quasi-condensate" is
in fact the vdBLP-GC. Here we generalise these results to the highly anisotropic $3D$-cuboid with
anisotropy in
\textit{one-dimension}, which is a model for infinite squared beams or cylinders, and "{cigar}"
type traps. \\
Second, we show that for the slab geometry with \textit{exponential} growing (for $\alpha > 0$ and
$L\rightarrow \infty$)
of two edges, $L_1=L_2=L e^{\alpha L}$, $L_3=L$,  of the anisotropic boxes:
$\Lambda=L_1\times L_2\times L_3\in\mathbb{R}^3$, there is a \textit{second} critical density
$\rho_{m}(\beta):= \rho_{c}(\beta)+ 2 \alpha / \lambda_{\beta}^{2} \geq \rho_{c}(\beta)$ such that
the vdBLP-GC
changes its properties when $\rho > \rho_{m}(\beta)$. This surprising behaviour of the BEC for the
PBG was discovered
by van den Berg \cite{vdB}, developed in \cite{vdBLL}, and then in \cite{GP},\cite{Pa} for the
spin-wave condensation.

Notice that the \textit{exponential} anisotropy is not a very common concept for the experimental
implementations. Therefore, it appeals for a re-examination of the standard vdBLP-GC concept in Casimir boxes
\cite{Beau} and the corresponding version of the Bogoliubov-Hohenberg theorem \cite{MuHoLa}.

Our original observation concerns the coexistence of two types of the vdBLP-GC for $\rho > \rho_{m}(\beta)$
(or for corresponding temperatures $T < T_m (\rho)$ for a fixed density) and the analysis of the coherence
length (ODLRO) in this anisotropic geometry. We extend also our observation to obtain another new result
proving the existence of the \textit{second} critical density in the {squared beam} and in the
"{cigar}" type traps for {exponentially weak} harmonic potential confinement in \textit{one} direction.
We use these results to calculate the temperature dependence of the vdBLP-GC particle density for the
case of two critical densities, $\rho_{m}(\beta) {>} \rho_{c}(\beta)$ and to apply the recent scaling
approach \cite{Beau} to the ODLRO asymptotic in this case.

\noindent 2. It is known that all kinds of BEC in the PBG  are defined by the \textit{limiting spectrum}
of the
one-particle Hamiltonian $T_{\Lambda}^{(N=1)}=-{\hbar^2}\Delta/({2m})$, when cuboid
$\Lambda \uparrow {\mathbb{R}}^3$.
In this paper we make this operator self-adjoint by fixing the \textit{Dirichlet} boundary conditions
on $\partial \Lambda$,
although our results are valid for all \textit{non-attractive} boundary conditions. Then the spectrum
is the set
\begin{equation}\label{spectre}
\{\varepsilon_{s} = \frac{\hbar^2}{2m}\sum_{j=1}^3 (\pi s_j/L_j)^2\}_{s_j\in \mathbb{N}}
\end{equation}
and $\{\phi_{s,\Lambda}(x)= \prod_{j=1}^3 \sqrt{2/L_j} \sin (\pi s_j x_j /L_j)\}_{s_j\in \mathbb{N}}$
are the eigenfunctions. Here $\mathbb{N}$ is the set of the natural numbers and
$s=(s_1,s_2,s_3)\in \mathbb{N}^3$
is the multi-index.

In the grand-canonical ensemble $(T,V,\mu)$, here $V=L_1 L_2 L_3$ is the volume of $\Lambda$, the mean
occupation
number of the state $\phi_{s,\Lambda}$ is $N_s(\beta,\mu)= (e^{\beta(\varepsilon_{s}-\mu)}-1)^{-1}$, where
$\mu<\inf_s \varepsilon_{s,\Lambda}$. Then for the fixed total particle density $\rho$ the
corresponding value of the
chemical potential $\mu_\Lambda (\beta,\rho)$ is a unique solution of the equation
$\rho = \sum_{s\in {\mathbb{N}}^3} N_s(\beta,\mu)/V =: N_{\Lambda}(\beta,\mu)/V$. Independent of the way
$\Lambda \uparrow {\mathbb{R}}^3$, one gets the limit
$\rho(\beta,\mu)=\lim_{V\rightarrow\infty}N_{\Lambda}(\beta,\mu)/V$,
which is the total particle density for $\mu \leq \lim_{V\rightarrow\infty}\inf_s \varepsilon_{s}=0$.
Since $\rho_c (\beta):=\sup_{\mu\leq 0}\rho(\beta,\mu)= \rho(\beta,\mu =0)< \infty$, it is the
(\textit{first}) critical
density for the $3D$ PBG: $\rho_c (\beta) = {\zeta(3/2)}/{\lambda_{\beta}^{3}}$. Here $\zeta(s)$ is
the Riemann
$\zeta$-function and $\lambda_{\beta}:=\hbar\sqrt{{2\pi \beta}/{m}}$ is the de Broglie thermal length.

\noindent 3. For $\Lambda= L e^{\alpha L}\times L e^{\alpha L}\times L$ one gets (\cite{vdB, vdBLL})
that for any $\mu\leq 0$ the limit of Darboux-Riemann sums
\begin{equation}\label{DR-n1n2}
\lim_{L\rightarrow\infty}\sum_{s \neq (s_1,s_2,1)} \frac{N_s(\beta,\mu)}{V_L} =
\frac{1}{(2 \pi)^3}\int_{\mathbb{R}^3}\frac{d^3 k}{e^{\beta(\hbar^2 k^2/2m-\mu)}-1} \ .
\end{equation}
We denote by $\mu_{L}(\beta,\rho):= \varepsilon_{(1,1,1)} - \Delta_{L}(\beta,\rho)$,
where $\Delta_{L}(\beta,\rho)\geq 0$ is a unique the solution of the equation:
\begin{equation}\label{Eq1}
\rho = \sum_{s = (s_1,s_2,1)} \frac{N_s(\beta,\mu)}{V_L} +
\sum_{s \neq (s_1,s_2,1)} \frac{N_s(\beta,\mu)}{V_L} .
\end{equation}
Since by (\ref{DR-n1n2}): $\lim_{L\rightarrow\infty}\sum_{s \neq (s_1,s_2,1)}{N_s(\beta,\mu=0)}/{V_L}=
\rho_c (\beta)$, for $\rho > \rho_c (\beta)$ the limit $L\rightarrow\infty$ of the first sum in
(\ref{Eq1}) is
equal to
\begin{eqnarray}\label{Eq2}
&&\lim_{L\rightarrow\infty}\sum_{s =(s_1,s_2,1)} \frac{N_s(\beta,\mu)}{V_L} =  \\
&&\lim_{L\rightarrow\infty} \frac{1}{L} \frac{1}{(2 \pi)^2}\int_{\mathbb{R}^2}\frac{d^2k}
{e^{\beta(\hbar^2 k^2/2m + \Delta_{L}(\beta,\rho))}-1} = \nonumber \\
&&\lim_{L\rightarrow\infty} - \frac{1}{\lambda_{\beta}^{2} L} \ln [\beta \Delta_{L}(\beta,\rho)] =
\rho - \rho_c (\beta).\nonumber
\end{eqnarray}
This implies the asymptotics:
\begin{equation}\label{ro<roM}
\Delta_{L}(\beta,\rho) = \frac{1}{\beta} \  e^{- \lambda_{\beta}^{2} (\rho - \rho_c (\beta)) L } +
\ldots \ .
\end{equation}

Notice that representation of the limit (\ref{Eq2}) by the integral (see (\ref{spectre})) is valid
only when
$\lambda_{\beta}^{2}(\rho - \rho_c (\beta)) < 2 \alpha$. For $\rho$ larger than the
\textit{second} critical density: $\rho_m(\beta):= \rho_c (\beta) + 2 \alpha / \lambda_{\beta}^{2}$ the
correction $\Delta_{L}(\beta,\rho)$ must converge to zero faster than $e^{- 2\alpha L }$. Now to keep the
difference $\rho - \rho_m (\beta) > 0$ we have to return back to the original sum representation
(\ref{Eq1}) and
(as for the standard BEC) to take into account the impact of the ground state occupation density
\textit{together}
with a saturated \textit{non-ground state} (i.e. generalised) condensation $\rho_m(\beta)- \rho_c (\beta)$
as in (\ref{Eq2}). For this case the asymptotics of $\Delta_{L}(\beta,\rho > \rho_m(\beta))$ is
completely different than (\ref{ro<roM}) and it is equal to
$\Delta_{L}(\beta,\rho) = [\beta (\rho - \rho_m(\beta))V_L]^{-1}$. Since $V_L= L^3 e^{2\alpha L}$, we
obtain:
\begin{eqnarray}\label{Eq3}
&&\lim_{L\rightarrow\infty}\sum_{s =(s_1 > 1,s_2 > 1, 1)} \frac{N_s(\beta,\mu)}{V_L}= \\
&&\lim_{L\rightarrow\infty} - \frac{1}{\lambda_{\beta}^{2} L} \ln [\beta \Delta_{L}(\beta,\rho)] =
2 \alpha / \lambda_{\beta}^{2} = \nonumber \\
&&\rho_m(\beta)- \rho_c (\beta), \nonumber
\end{eqnarray}
and the ground-state term gives the \textit{macroscopic} occupation:
\begin{equation}\label{ro>roM}
\rho - \rho_m (\beta)= \lim_{L\rightarrow\infty}
\frac{1}{V_L}\frac{1}{e^{\beta(\varepsilon_{(1,1,1)} - \mu_{L}(\beta,\rho))}-1} \ .
\end{equation}
\begin{figure}
\begin{center}
  \includegraphics[scale=0.7]{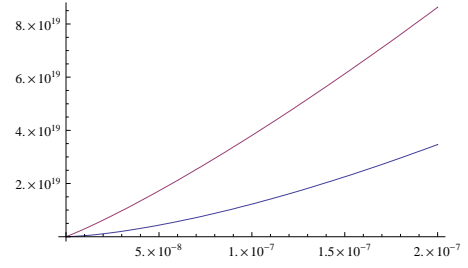}\\
  \caption{{For the slab geometry}, the blue curve $\rho_c (1/(k_B T))$ is the first critical line for the
  BEC transition as a function of $T$,
  the red curve $\rho_m (1/(k_B T)) = \rho_c (1/(k_B T)) + 2\alpha /\lambda^{2}_{\beta}$ is the
  second critical line.
  Notice that above the red curve there is {a \emph{coexistence} between "quasi-condensate"
  (vdBLP-GC of type III)
  and the \emph{conventional} condensate in the ground state (vdBLP-GC of type I), between two
  curve there is only
  "quasi-condensates" phase and below the blue curve there is no condensate.}}\label{Diagram (n-T)}
\end{center}
\end{figure}
Notice that for $\rho_c(\beta)< \rho< \rho_m (\beta)$ we obtain the vdBLP-GC
(of the \textit{type} III), i.e. \textit{none} of the single-particle states are \textit{macroscopically}
occupied,
since by virtue of (\ref{spectre}) and (\ref{ro<roM}) for any $s$ one has:
\begin{equation}\label{typeIII}
\rho_s (\beta,\rho):= \lim_{L\rightarrow\infty}
\frac{1}{V_L}\frac{1}{e^{\beta(\varepsilon_{s} - \mu_{L}(\beta,\rho))}-1} = 0 \ .
\end{equation}
On the other hand, the asymptotics $\Delta_{L}(\beta,\rho> \rho_m(\beta)) =
[\beta (\rho - \rho_m(\beta))V_L]^{-1}$
implies
\begin{equation}\label{typeI-typeIII}
\rho_{s\neq (1,1,1)} (\beta,\rho):= \lim_{L\rightarrow\infty}
\frac{1}{V_L}\frac{1}{e^{\beta(\varepsilon_{s} - \mu_{L}(\beta,\rho))}-1} = 0 \ ,
\end{equation}
i.e. for $\rho > \rho_m(\beta)$ there is a \textit{coexistence} of the \textit{saturated} type
III vdBLP-GC, with the
constant density (\ref{Eq3}), and the standard BEC (i.e. the \textit{type} I vdBLP-GC) in the
single state (\ref{ro>roM}).

\noindent 4. It is curious to note that neither Casimir shaped boxes \cite{vdBLP}, nor the van den Berg
boxes $\Lambda= L e^{\alpha L}\times L \times L$, with one-dimensional anisotropy do not produce the
\textit{second}
critical density $\rho_m(\beta)\neq\rho_c (\beta)$. To model infinite squared beams with BEC
transitions at two
critical densities we propose the one-particle Hamiltonian:
$T_{\Lambda}^{(N=1)}= -{\hbar^2}\Delta/({2m}) + m \omega^{2}_{1} x_1 ^2/2$, with \textit{harmonic trap}
in direction
$x_1$ and, e.g., Dirichlet boundary conditions in directions $x_2,x_3$. Then the spectrum is the set
\begin{equation}\label{spectre-harm}
\{\epsilon_{s} := \hbar \omega_{1}(s_1 + {1}/{2}) +
\frac{\hbar^2}{2m}\sum_{j=2}^3 (\pi s_j/L_j)^2\}_{s \in \mathbb{N}}.
\end{equation}
Here multi-index $s=(s_1,s_2,s_3)\in (\mathbb{N}\cup\{0\})\times\mathbb{N}^2$, and the ground-state
energy is
$\epsilon_{(0,1,1)}$. Then for $\mu_{L}(\beta,{\varrho}):= \epsilon_{(0,1,1)} -
\Delta_{L}(\beta,{\varrho})$, the
value of $\Delta_{L}(\beta,{\varrho})\geq 0$, is a solution of the equation:
\begin{equation}\label{Eq1-harm}
{\varrho}:= \sum_{s = (s_1,1,1)} \omega_{1} \frac{N_s(\beta,\mu)}{L_2 L_3} +
\sum_{s \neq (s_1,1,1)} \omega_{1} \frac{N_s(\beta,\mu)}{L_2 L_3} ,
\end{equation}
where $N_s(\beta,\mu)=(e^{\beta(\epsilon_{s}-\mu)}-1)^{-1}$.

Let $\omega_{1}:= \hbar/(m L_{1}^2)$ and $L_2 = L_3 = L$. Here $L_{1}$ is the harmonic-trap
characteristic size in
direction $x_1$. Then for any $s_1\geq 0$ and $\mu\leq 0$
\begin{eqnarray}
&&\varrho (\beta,\mu):=
\lim_{L_1,L\rightarrow\infty}\sum_{s \neq (s_1,1,1)} \omega_{1} \frac{N_s(\beta,\mu)}{L_2 L_3} =
\nonumber \\
&&\frac{1}{(2 \pi)^2}\int_{0}^{\infty} dp\int_{\mathbb{R}^2}\frac{d^2 k}{e^{\beta(\hbar p +
\hbar^2 k^2/2m-\mu)}-1} \ . \label{DR-n1}
\end{eqnarray}
Therefore, the \textit{first} critical density is \textit{finite}: $\varrho_{c}(\beta):=
\sup_{\mu\leq 0}\varrho(\beta,\mu)
=\varrho (\beta,\mu =0) < \infty$. If $\varrho> \varrho_{c}(\beta)$, then the limit
$L\rightarrow\infty$ of the
first sum in (\ref{Eq1-harm}) is
\begin{eqnarray}\label{Eq2-harm}
&&\lim_{L_1,L\rightarrow\infty}\sum_{s =(s_1,1,1)} \omega_{1}\frac{N_s(\beta,\mu_L)}{L_2 L_3} =  \\
&&\lim_{L\rightarrow\infty} \frac{1}{L^2} \int_{0}^{\infty}\frac{d p}{e^{\beta(\hbar p +
\Delta_{L}(\beta,{\varrho}))}-1}
= \nonumber \\
&&\lim_{L\rightarrow\infty} - \frac{1}{\hbar\beta L^2} \ln [\beta \Delta_{L}(\beta,{\varrho})] =
{\varrho} - \varrho_c (\beta).\nonumber
\end{eqnarray}
This means that the asymptotics of $\Delta_{L}(\beta,\rho)$ is:
\begin{equation}\label{ro<roM-harm}
\Delta_{L}(\beta,{\varrho}) = \frac{1}{\beta} \  e^{- \hbar \beta ({\varrho} -
\varrho_c (\beta)) L^2} + \ldots \ .
\end{equation}

Let $L_1:= L e^{\gamma L^2}$, for $\gamma > 0$. Then, similar to our arguments in 2., the
representation of the limit
(\ref{Eq2-harm}) by the integral is valid for $\hbar \beta (\varrho - \varrho_c (\beta)) < 2 \gamma$.
For ${\varrho}$ larger
than the \textit{second} critical density: $\varrho_m(\beta):= \varrho_c (\beta) +
2 \gamma /(\hbar \beta)$ the
chemical potential correction (\ref{ro<roM-harm}) must converge to zero faster than $e^{- 2 \gamma L^2}$.
By the same line of reasoning as in 2., to keep the difference ${\varrho} - \varrho_m (\beta) > 0$ we
have to use
the original sum representation (\ref{Eq1-harm}) and to take into account the input due to the
ground state occupation density \textit{together} with a saturated \textit{non-ground state}
condensation $\varrho_m(\beta)- \varrho_c (\beta)$ (\ref{Eq2-harm}). The asymptotics of
$\Delta_{L}(\beta,{\varrho} > \varrho_m(\beta))$ is than equal to $\Delta_{L}(\beta,{\varrho}) =
[\beta m ({\varrho} - \varrho_m(\beta)) L^4 e^{2 \gamma L^2}/\hbar]^{-1}$. Hence,
\begin{eqnarray}\label{Eq3-harm}
&&\lim_{L\rightarrow\infty}\sum_{s =(s_1 > 0,1,1)} \frac{\hbar}{m} \
\frac{N_s(\beta,\mu_L)}{L^4 e^{2 \gamma L^2}}= \\
&&\lim_{L\rightarrow\infty} - \frac{1}{\hbar\beta L^2} \ln [\beta \Delta_{L}(\beta,{\varrho})] =
\frac{2 \gamma}{\hbar\beta} = \varrho_m(\beta)- \varrho_c (\beta), \nonumber
\end{eqnarray}
and the ground-state term gives the \textit{macroscopic} occupation:
\begin{equation}\label{ro>roM-harm}
{\varrho} - \varrho_m (\beta)= \lim_{L\rightarrow\infty}
\frac{\hbar}{m L^4 e^{2 \gamma L^2}}\frac{1}{e^{\beta(\epsilon_{(0,1,1),L} -
\mu_{L}(\beta,{\varrho}))}-1} \ .
\end{equation}

With this choice of boundary conditions and the one-dimensional anisotropic trap our model of the
infinite squared beams manifests the BEC with two critical densities. Again for
$\varrho_c(\beta)< {\varrho}< \varrho_m (\beta)$
we obtain the \textit{type} III vdBLP-GC, i.e., \textit{none} of the single-particle states are
\textit{macroscopically}
occupied:
\begin{equation}\label{typeIII-harm}
\varrho_s (\beta,{\varrho}):= \lim_{L\rightarrow\infty}
\frac{\hbar}{m L^4 e^{2 \gamma L^2}}\frac{1}{e^{\beta(\epsilon_{s} - \mu_{L}(\beta,{\varrho}))}-1} = 0 \ .
\end{equation}
When $\varrho_m(\beta)< {\varrho}$ there is a coexistence of the \textit{type} III vdBLP-GC,
with the constant density
(\ref{Eq3-harm}), and the standard \textit{type} I vdBLP-GC  in the single state (\ref{ro>roM-harm}),
since
\begin{equation}\label{typeI-typeIII-harm}
\varrho_{s\neq (0,1,1)} (\beta,{\varrho}):= \lim_{L\rightarrow\infty}
\frac{\hbar}{m L^4 e^{2 \gamma L^2}}\frac{1}{e^{\beta(\epsilon_{s} - \mu_{L}(\beta,{\varrho}))}-1} = 0 \ .
\end{equation}

Finally, it is instructive to study a "cigar"-type geometry ensured by the anisotropic harmonic trap:
\begin{equation}\label{1D-harm-trap}
T_{\Lambda}^{(N=1)}= -{\hbar^2}\Delta/({2m}) + \sum_{1\leq j \leq3} m \omega^{2}_{j} x_j ^2/2 \ .
\end{equation}
with $\omega_{1} = \hbar/(m L_{1}^2) , \omega_{2}=\omega_{3}=\hbar/(m L^2)$. Here $L_{1}, L_{2}=L_{3}=L$
are the \textit{characteristic} sizes of the trap in three directions and
$\eta_{s} = \sum_{1\leq j \leq3}\hbar \omega_{j}(s_j + {1}/{2})$ is the corresponding one-particle spectrum.
Then the same reasoning as in (\ref{DR-n1}),(\ref{Eq2-harm}), yields for
$\mu_{L}(\beta,{n}):= \eta_{(0,0,0)} - \Delta_{L}(\beta,{n})$ and auxiliary dimensionality factor
$\kappa >0$:
\begin{eqnarray}\label{Eq3-harm}
&&\lim_{L_1,L\rightarrow\infty}\sum_{s =(s_1,0,0)} \kappa^3 \omega_{1}\omega_{2}\omega_{3}
{N_s(\beta,\mu_L)} =  \\
&&\lim_{L\rightarrow\infty} - \frac{\kappa^3 \hbar}{\beta(m L^2)^2} \ln [\beta \Delta_{L}(\beta,{n})] =
{n} - n_c (\beta).\nonumber
\end{eqnarray}
Here the \textit{finite} critical density $n_c (\beta):=n(\beta, \mu =0)$ is defined similarly
to (\ref{DR-n1}),
where the particle density is
\begin{eqnarray}
&&n(\beta,\mu):=
\lim_{L_1,L\rightarrow\infty}\sum_{s \neq (s_1,0,0)} \kappa^3 \omega_{1} \omega_{2} \omega_{2}
{N_s(\beta,\mu)} = \nonumber \\
&&\int_{{\mathbb{R}^3}_+}\frac{\kappa^3 d{\omega}_1 d{\omega}_2 d{\omega}_3}
{e^{\beta [\hbar({\omega}_1 + {\omega}_2 + {\omega}_3)-\mu]}-1} \ . \label{DR-n1-harm}
\end{eqnarray}
Equation (\ref{Eq3-harm}) implies for $\Delta_{L}(\beta,{n})$ the same asymptotics as in (\ref{Eq2-harm}):
\begin{equation}\label{ro<roM-harm-cigar}
\Delta_{L}(\beta,{n}) = \frac{1}{\beta} \  e^{- \beta ({n} - n_c (\beta))m^2 L^4/(\hbar \kappa^3)} +
\ldots \ .
\end{equation}
If we choose $L_1:= L e^{\widehat{\gamma} L^4}$, for $\widehat{\gamma} > 0$, then the \textit{second}
critical density
$n_m(\beta):= n_c (\beta) + (\widehat{\gamma} \hbar \kappa^3)/(\beta m^2)$. For
$n_c(\beta)< {n}< n_m (\beta)$
we obtain the \textit{type} III vdBLP-GC, i.e., \textit{none} of the single-particle states are
\textit{macroscopically}
occupied:
\begin{equation}\label{typeIII-harm-cigar}
n_s (\beta,n):= \lim_{L\rightarrow\infty}
\frac{\kappa^3 \, \omega_{1}\omega_{2}\omega_{3}}{e^{\beta(\eta_{s} - \mu_{L}(\beta,{n}))}-1} = 0 \ .
\end{equation}
Although for $n_m(\beta)< n$ there is a coexistence of the \textit{type} III vdBLP-GC, with the
constant density
$n_m (\beta)-n_c (\beta)$, and the standard \textit{type} I vdBLP-GC  in the ground-state:
\begin{equation}\label{typeI-typeIII-harm-cigar}
{n} - n_m (\beta)= \lim_{L\rightarrow\infty}
\frac{\kappa^3 \, \omega_{1}\omega_{2}\omega_{3}}{e^{\beta(\eta_{(0,0,0)} - \mu_{L}(\beta,{n}))}-1} \ .
\end{equation}

\noindent 5. In experiments with BEC, it is important to know the critical temperatures associated with
corresponding
critical densities. The \textit{first} critical temperatures: $T_c (\rho)$, $\widetilde{T}_c (\rho)$ or
$\widehat{T}_c (\rho)$ are well-known. For a given density $\rho$ they verify the identities:
\begin{equation}\label{Tc}
\rho = \rho_c (\beta_c (\rho)) \ , \ \varrho = \varrho_c (\widetilde{\beta}_c (\varrho)) \ , \
n = n_c (\widehat{\beta}_c (n)) \ ,
\end{equation}
respectively for our models of slabs, squared beams or "cigars". Since definition of the critical
densities yield the representations: $\rho_c (\beta) =: T^{3/2} \, I_{sl}$,
$\varrho_c (\beta) =: T^{2}\, I_{bl}$, $n_c (\beta) =: T^{3}\, I_{cg}$, the expressions for the
\textit{second} critical
densities one gets the following relations between the \textit{first} and the \textit{second}
critical temperatures:
\begin{eqnarray*}
T_{m}^{3/2}(\rho)+{{\tau}^{1/2}}\ T_{m}(\rho)&=&T_{c}^{3/2}(\rho) \  \ {\rm{(slab)}} \ , \\
\widetilde{T}_{m}^{2}({\varrho})+{\widetilde{\tau}}\ \widetilde{T}_{m}({\varrho})&=&
\widetilde{T}_{c}^{2}({\varrho}) \  \ \ \ {\rm{(beam)}} \ , \\
\widehat{T}_{m}^{3}({n})+{\widehat{\tau}}^2\ \widehat{T}_{m}({n})&=&
\widehat{T}_{c}^{3}({n}) \ \ \ \ {\rm{(cigar)}} \ .
\end{eqnarray*}
Here $\tau = [\alpha m k_B/(\pi \hbar^2 I_{sl})]^2$, $\widetilde{\tau} =  2 \gamma k_B/(\hbar I_{bl})$
and $\widehat{\tau} = [(\widehat{\gamma} \hbar \kappa^3 k_B) /(m^2 I_{cg})]^{1/2}$ are
"effective" temperatures related to the corresponding geometrical shapes.
\begin{figure}
\begin{center}
  \includegraphics[scale=0.7]{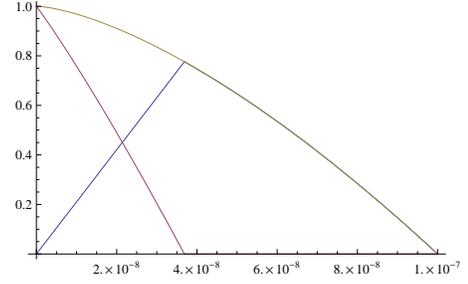}\\
  \caption{The first (blue fit) , The second (pink fit) and the total (green fit) condensate
  fractions as a
  function of the temperature
  for $^{87}Rb$ atoms {in the slab geometry} with $T_{c_1}=10^{-7}K$ and
  $\tau=4.43\times10^{-7}$.}\label{FractionsBEC}
\end{center}
\end{figure}
Notice that the \textit{second} critical temperature modifies the usual law for the condensate fractions
temperature dependence, since now the total condensate density is
$\rho - \rho_{c} (\beta):= \rho_{0}(\beta) = \rho_{0c}(\beta) + \rho_{0m}(\beta)$. Here $\rho_{0m}(\beta):=
(\rho -\rho_{m} (\beta)) \ \theta (\rho - \rho_{m} (\beta))$.

For example, in the case of the slab geometry the \textit{type} III vdBLP-GC (i.e. the "quasi-condensate")
$\rho_{0c}(\beta)$ behaves for a given $\rho$ like (see Fig.2.)
\begin{equation}\label{BEC-III}
\frac{\rho_{0c}(\beta)}{\rho} =
\left\{ \begin{array}{ll}
1 - (T/T_{c})^{3/2} \ , & \textrm{~~~~~$T_{m} \leq T \leq T_{c}$ }, \\
{\sqrt{\tau} \ T}/{T_{c}^{3/2}} \ \ \ \ ,  & \textrm{~~~~~$ T \leq T_{m}$ }.
\end{array} \right.
\end{equation}
Similarly, for the \textit{type} I vdBLP-GC in the ground state $\rho_{0m}(\beta)$
(i.e. the conventional BEC)
we obtain:
\begin{equation}\label{BEC-I}
\frac{\rho_{0m}(\beta)}{\rho} =
\left\{ \begin{array}{ll}
0 \ , & \textrm{$T_{m} \leq T \leq T_{c}$}, \\
1 - (T/T_{c})^{3/2}(1 + \sqrt{{\tau}/{T}}),  & \textrm{$ T \leq T_{m}$},
\end{array} \right.
\end{equation}
see Fig.2. The total condensate density $\rho_{0}(\beta):=\rho_{0c}(\beta)+\rho_{0m}(\beta)$ is
the result of
\textit{coexistence} of both of them: it gives the standard PBG expression $\rho_{0}(\beta)/\rho =
1 - (T/T_{c})^{3/2}$.

For the "cigars" geometry case the temperature dependence of the "quasi-condensate" $r_{0c}(\beta)$ is
\begin{equation}\label{BEC-III-sg}
\frac{n_{0c}(\beta)}{{n}} =
\left\{ \begin{array}{ll}
1 - (T/\widehat{T}_{c})^{3} \ , & \textrm{~~~~~$\widehat{T}_{m} \leq T \leq \widehat{T}_{c}$ }, \\
{{\widehat{\tau}}^2 \ T}/{\widehat{T}_{c}^{3}} \ \ \ \ \ \ ,  & \textrm{~~~~~$ T \leq \widehat{T}_{m}$ }.
\end{array} \right.
\end{equation}
The corresponding ground state conventional BEC behaves as
\begin{equation}\label{BEC-I-sg}
\frac{n_{0m}(\beta)}{{n}} =
\left\{\begin{array}{ll}
0 \ , & \textrm{${\widehat{T}}_{m} \leq T \leq {\widehat{T}}_{c}$}, \\
1 - (T/\widehat{T}_{c})^{3}(1 +{\widehat{\tau}}^2 /{T^{2}}),  & \textrm{$ T \leq {\widehat{T}}_{m}$},
\end{array} \right.
\end{equation}
and again for the two coexisting condensates one gets
${n} - n_{c} (\beta):= n_{0}(\beta)= n_{0c}(\beta)+n_{0m}(\beta) = (1 - (T/T_{c})^{3/2}) {n}$.

{Notice that for a given density the difference between two critical temperatures
{for the slab geometry} can be
calculated explicitly:
\begin{equation}\label{diff-temp}
(T_{c} - T_{m})/T_{c} = g(\rho_{\alpha}/\rho) \ ,
\end{equation}
where $\rho_{\alpha}:= 8 \alpha^3/\zeta(3/2)^{2}$ and $g(x)$ is an explicit algebraic function.
For illustration consider a
\textit{quasi}-$2D$ PBG model of $^{87}Rb$ atoms in trap with characteristic sizes
$L_1= L_2 = 100 \mu m$, $L= 1\mu m$
and with typical critical temperature $T_{c} = 10^{-7} K$. The \textit{anisotropy} parameter is
$\alpha=(1/L)\ln(L_1/L)=4,6 \cdot 10^{6} m^{-1}$.  Then for ${\tau} = 4,4 \cdot 10^{-7} K$ we find
$T_{m}=3,7 \cdot 10^{-8} K$ and $(T_{c}-T_{m})/T_{c}=0,63 $.}

\noindent 6. Another physical observable to characterise this second critical temperature
is the condensate coherence length or the global spacial  particle density distribution.
The usual criterion is the ODLRO, which is going back to Penrose and Onsager \cite{PO}. For a fixed
particle density $\rho$ it is defined by the kernel:
\begin{equation}\label{2-point-corr}
K(x,y):=\lim_{L\rightarrow\infty} K_{\Lambda}(x,y)= \lim_{L\rightarrow\infty}
\sum_{s} \frac{\overline{\phi}_{s,\Lambda}(x) \phi_{s,\Lambda}(y)}{e^{\beta(\varepsilon_s -
\mu_{L}(\beta,\rho))} - 1} \ .
\end{equation}
The limiting diagonal function $\rho(x):=K(x,x)$ is \textit{local} $x$-independent particle density.

To detect a trace of the geometry (or the second critical temperature) impact on the spatial
density distribution we follow a recent scaling approach to the generalised BEC developed in \cite{Beau}
(see also \cite{vdBLP},\cite{vdBLL}) and introduce a \textit{scaled global} particle density:
\begin{equation}\label{scaled-dens}
\xi_{L}(u) :=
\sum_{s} \frac{|{\phi}_{s,\Lambda}(L_1 u_1,L_2 u_2,L_3 u_3)|^2}
{e^{\beta(\varepsilon_s - \mu)} - 1} \ ,
\end{equation}
with the scaled distances $\{u_j = x_j/L_j \in [0,1] \}_{j=1,2,3}$.

For a given $\rho$ the scaled density (\ref{scaled-dens}) in the slab geometry is
\begin{equation}\label{sl-glob-dens}
\xi_{\rho,L}^{sl}(u) := \sum_{s} \frac{1}{e^{\beta(\varepsilon_s - \mu_{L}(\beta,\rho))} - 1}
\prod_{j=1}^{d=3} \frac{2}{L_j} [\sin(\pi s_j u_j)]^2 .
\end{equation}
Since $2 [\sin(\pi s_j u_j)]^2 = 1 - \cos \{(2 \pi s_j/L_j)u_j L_j\}$ and
$\lim_{L\rightarrow\infty}\mu_{L}(\beta,\rho<\rho_{c}(\beta))<0$, by the
Riemann-Lebesgue lemma we obtain that $\lim_{L\rightarrow\infty} \xi_{\rho,\Lambda}^{sl}(u) = \rho$
for any $u \in (0,1)^3$. If $\rho>\rho_{c}(\beta)$, one has to proceed as in (\ref{Eq1})-(\ref{ro<roM}).
Then for any $u \in (0,1)^3$:
\begin{eqnarray}
&& \lim_{L\rightarrow\infty} \sum_{s =(s_1,s_2,1)} \frac{1}{e^{\beta(\varepsilon_s -
\mu_{L}(\beta,\rho))} - 1}
\prod_{j=1}^{d=3} \frac{2}{L_j} [\sin(\pi s_j u_j)]^2 \nonumber \\
&&=\lim_{L\rightarrow\infty} \frac{2[\sin(\pi u_3)]^2}{(2 \pi)^2 L}\int_{\mathbb{R}^2}
\frac{\prod_{j=1}^{2}(1-\cos(2 k_j u_j L_j)  d^2k}
{e^{\beta(\hbar^2 k^2/2m + \Delta_{L}(\beta,\rho))}-1}\nonumber \\
&& =(\rho - \rho_c (\beta)) \ 2[\sin(\pi u_3)]^2 \ , \label{sl1-rho>}\\
&& \lim_{L\rightarrow\infty} \sum_{s \neq(s_1,s_2,1)} \frac{1}{e^{\beta(\varepsilon_s -
\mu_{L}(\beta,\rho))} - 1}
\prod_{j=1}^{d=3} \frac{2}{L_j} [\sin(\pi s_j u_j)]^2 \nonumber \\
&& = \rho_c (\beta)) \ . \label{sl2-rho>}
\end{eqnarray}
Then the limit of (\ref{sl-glob-dens}) is equal to
\begin{equation}\label{sl-rho<rhoM}
\xi_{\rho}^{sl}(u) =
(\rho - \rho_c (\beta)) \ 2[\sin(\pi u_3)]^2 + \rho_c (\beta) \ .
\end{equation}
It manifests a \textit{space anisotropy} of the type III vdBLP-GC for $\rho_c (\beta) < \rho <
\rho_m (\beta)$
in direction $u_3$.

For $\rho > \rho_m (\beta)$ one has to use representation (\ref{Eq1}) and asymptotics (\ref{Eq3}),
(\ref{ro>roM}).
Then following the arguments developed above we obtain
\begin{eqnarray}
&&\xi_{\rho}^{sl}(u) = (\rho - \rho_m (\beta)) \prod_{j=1}^3 \ 2[\sin(\pi u_j)]^2 + \nonumber \\
&&(\rho_m (\beta)-\rho_c (\beta)) \ 2[\sin(\pi u_3)]^2  + \rho_c (\beta) \ . \label{sl-rho>rhoM}
\end{eqnarray}
So, the anisotropy of the space particle distribution is still in direction $u_3$ due to the
type III vdBLP-GC.

It is instructive to compare this anisotropy with a \textit{coherence length} analysis within the scaling
approach \cite{Beau} to the BEC space distribution. To this end let us center the box $\Lambda$ at the
origin of coordinates: $x_j = \tilde{x}_j + L_j/2$ and $y_j = \tilde{y}_j + L_j/2$. Then the ODLRO kernel
(\ref{2-point-corr}) is:
\begin{equation}\label{Ker}
K_{\Lambda}(\tilde{x},\tilde{y}) = \sum_{l=1}^{\infty} e^{{l} \beta \mu_{L}(\beta,\rho)} \
R_{{l}}^{(2)} \ R_{{l}}^{(1)} \ ,
\end{equation}
where after the shift of coordinates and using (\ref{spectre}) we put
\begin{eqnarray} \label{Ker-2}
&&R_{{l}}^{(2)}(\tilde{x}^{(2)},\tilde{y}^{(2)})= \\
&& \sum_{s = (s_1,s_2)} e^{- {l} \beta \varepsilon_{s_1,s_2}} \
\overline{\phi}_{s_1,s_2,\Lambda}(\tilde{x}_1,\tilde{x}_2) \
{\phi}_{s_1,s_2,\Lambda}(\tilde{y}_1,\tilde{y}_2) \nonumber \\
&& R_{s}^{(1)}(\tilde{x}_3,\tilde{y}_3)=
\sum_{s = (s_3)} e^{- l \beta \varepsilon_{s_3}} \
\sqrt{\frac{2}{L_3}}\sin(\frac{\pi s_3}{L_3} (\tilde{x}_3 + \frac{L_3}{2})) \nonumber \\
&& \times \sqrt{\frac{2}{L_3}}\sin(\frac{\pi s_3}{L_3} (\tilde{y}_3 + \frac{L_3}{2})) \ .
\label{Ker-1}
\end{eqnarray}

Similar to (\ref{Eq1}), for $\rho_c (\beta)<\rho <\rho_m (\beta)$  we must split the sum over
$s = (s_1,s_2,s_3)$ in (\ref{Ker}) into two parts. Since by the generalized Weyl theorem one gets:
\begin{equation}\label{lim-Ker-2}
\lim_{L\rightarrow\infty} R_{{l}}^{(2)}(\tilde{x}^{(2)},\tilde{y}^{(2)}) =
\frac{1}{{l} \lambda_{\beta}^{2}} \  e^{-\pi \|\tilde{x}^{(2)} - \tilde{y}^{(2)}\|^2/{l}
\lambda_{\beta}^{2}}
\nonumber \ ,
\end{equation}
by (\ref{Ker}) we obtain for the first part the representation:
\begin{eqnarray}\label{lim-Ker-3-1}
&&\lim_{L\rightarrow\infty} \sum_{{l}=1}^{\infty} e^{{l} \beta \mu_{L}(\beta,\rho)}
\sum_{s = (s_1,s_2,1)} e^{- {l} \beta \varepsilon_{s_1,s_2,1}} \times\\ \nonumber
&&\times\overline{\phi}_{s_1,s_2, 1\Lambda}(\tilde{x}) \ {\phi}_{s_1,s_2,1\Lambda}(\tilde{y}) = \\
\nonumber
&&\lim_{L\rightarrow\infty} \sum_{{l}=1}^{\infty} e^{-{l} \beta \Delta_{L}(\beta,\rho)}
\frac{1}{{l} \lambda_{\beta}^{2}} \  e^{-\pi \|\tilde{x}^{(2)} - \tilde{y}^{(2)}\|^2/{l}
\lambda_{\beta}^{2}}\times
\\ \nonumber
&&\times\frac{2}{L} \sin(\frac{\pi}{L}(\tilde{x}_3 + \frac{L}{2})) \sin(\frac{\pi}{L}(\tilde{y}_3 +
\frac{L}{2})) \ .
\end{eqnarray}
For the second part we apply the Weyl theorem for the 3-dimensional Green function:
\begin{eqnarray}\label{lim-Ker-3-3}
&&\lim_{L\rightarrow\infty} \sum_{{l}=1}^{\infty} e^{{l} \beta \mu_{L}(\beta,\rho)}
\sum_{s \neq (s_1,s_2, 1)} e^{- {l} \beta \varepsilon_{s}} \times \\ \nonumber
&&\times\overline{\phi}_{s,\Lambda}(\tilde{x}) \ {\phi}_{s,\Lambda}(\tilde{y}) =
\sum_{{l}=1}^{\infty} \frac{1}{{l} \lambda_{\beta}^{3}} \  e^{-\pi \|\tilde{x} - \tilde{y}\|^2/{l}
\lambda_{\beta}^{2}} \ .
\end{eqnarray}
If in (\ref{lim-Ker-3-1}) we change ${l} \rightarrow {l} \ \Delta_{L}(\beta,\rho)$, then it gets
the form of the
integral Darboux-Riemann sum, where $\|\tilde{x}^{(2)} - \tilde{y}^{(2)}\|^2$ is scaled as
$\|\tilde{x}^{(2)} - \tilde{y}^{(2)}\|^2 \ \Delta_{L}(\beta,\rho)$. Therefore, the
\textit{coherence length} $L_{ch}$
in direction perpendicular to $x_3$ is $L_{ch}(\beta,\rho)/L := 1/\sqrt{\Delta_{L}(\beta,\rho)}$.
A similar argument is valid for $\rho >\rho_m (\beta)$ with obvious modifications due to BEC
for $s=(1,1,1)$ (\ref{ro>roM}) and to another asymptotics (\ref{Eq3}) for $\Delta_{L}(\beta,\rho)$.
To compare the coherence length with the \textit{scale} $L_{1,2}= L e^{\alpha L}$, let us define the
critical exponent $\gamma(T,\rho)$ such that
$\lim_{L\rightarrow\infty} (L_{ch}(\beta,\rho)/L)({L_{1}}/{L})^{-\gamma(T,\rho)} =1$. Then we get:
\begin{eqnarray}
\gamma(T,\rho)&=& {\lambda_{\beta}^{2}}\ (\rho - \rho_{c}(\beta))/{2 \alpha} \ ,\
\rho_{c}(\beta)<\rho<\rho_{m}(\beta) \nonumber \\
&=& {\lambda_{\beta}^{2}} \ (\rho_{m}(\beta) - \rho_{c}(\beta))/{2 \alpha} \ ,\
\rho_{m}(\beta) \leq \rho \ . \label{gamma(rho)}
\end{eqnarray}
For a fixed density, taking into account (\ref{BEC-III}) we find the temperature dependence of
the exponent $\gamma(T):=\gamma(T,\rho)$, see Fig.3:
\begin{eqnarray}
\gamma(T)&=&\sqrt{{T}/{\tau}} \ \{({T_{c}}/{T})^{3/2}-1\} \ ,\ T_{m}<T<T_{c} \ , \nonumber \\
&=&1,\ T \leq T_{m} \ . \label{gamma(T)}
\end{eqnarray}
Notice that in the both cases the ODLRO kernel is anisotropic due to impact of the type III condensation
(\ref{lim-Ker-3-1}) in the states $s = (s_1,s_2, 1)$, whereas the other states give a symmetric part of
correlations (\ref{lim-Ker-3-3}), which includes a constant term $\rho_{c}(\beta)$.

\begin{figure}
\begin{center}
  \includegraphics[scale=0.7]{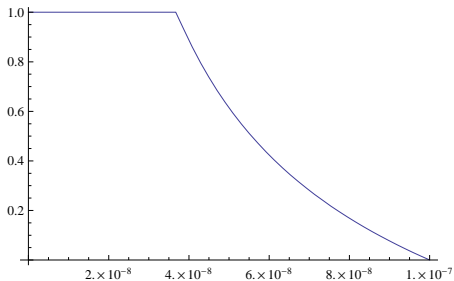}\\
  \caption{Exponent $\gamma(T)$ for evolution of the coherence length for the quasi-condensate
  with temperature
  corresponding to  $^{87}Rb$ atoms {in the slab geometry} with $T_{c}=10^{-7}K$ and
  $\tau=4.43\times10^{-7}K$}
  \label{ExposantLongCoherence}
\end{center}
\end{figure}

Numerically, for $L_1=L_2=100 \mu m,\ L_3=1 \mu m$ and  $T_{m} < T=0.75 T_{c}$
the coherence length of the condensate  is equal to $2.8\mu m\ll 100 \mu m$.
This decreasing of the \textit{coherence length} is experimentally observed in \cite{Gerbier}.

\noindent 7. In conclusion we add several remarks about a possible impact of particle interaction.
Since the "quasi-condensate" is observed in extremely anisotropic traps \cite{Gerbier}-\cite{Petrov1},
we think that the geometry of the vessels is predominant. So, the study of the PBG is able to catch the
phenomenon and so is relevant. Next, in this letter did not enter into details of the phase-fluctuations
\cite{Petrov2}, \cite{Gerbier},
although we suppose that for the vdBLP-GC it can be studied by switching different Bogoliubov
quasi-average sources in condensed modes. Finally, since a \textit{repulsive} interaction is able to
\textit{transform} the conventional one-mode BEC (\textit{type} I) into the vdBLP-GC of \textit{type} III,
\cite{MV}, \cite{BZ}, it is important to combine study of this interaction with the results already
obtained for interacting gases in \cite{Gerbier}-\cite{Petrov1} and in \cite{MuHoLa}.

The pioneer calculations of a crossover in a trapped $1D$ PBG are due to \cite{KvD}. It is similar to
the vdBLP-GC  in our exact calculations for the "cigars" geometry and it apparently persists for a
\textit{weakly} interacting Bose-gas as argued in \cite{Petrov1}. Although the ultimate aim is to
understand the relevance of these quasi-$1D$ calculations for the Lieb-Liniger \textit{exact} analysis of
a strongly interacting gas \cite{LL}. We return to these questions in our next papers.


\end{document}